\documentclass[%
 reprint,
 superscriptaddress,
nofootinbib,
 amsmath,amssymb,
 aps,
]{revtex4-2}

\usepackage{xcolor}
\usepackage{graphicx, cleveref, siunitx, subfigure} 
\usepackage{dcolumn}
\usepackage{bm}

\begin{document}

\preprint{APS/123-QED}

\title{Horn-array haloscope for volume-efficient broadband axion searches}

\author{Junu Jeong}
\email{jwpc0120@ibs.re.kr}
\affiliation{Center for Axion and Precision Physics Research (CAPP), Institute for Basic Science (IBS), Daejeon 34051, Republic of Korea}

\author{Sungwoo Youn}
\email{swyoun@ibs.re.kr}
\affiliation{Center for Axion and Precision Physics Research (CAPP), Institute for Basic Science (IBS), Daejeon 34051, Republic of Korea}

\author{Yannis K. Semertzidis}
\email{semertzidisy@gmail.com}
\affiliation{Center for Axion and Precision Physics Research (CAPP), Institute for Basic Science (IBS), Daejeon 34051, Republic of Korea}
\affiliation{Department of Physics, Korea Advanced Institute of Science and Technology (KAIST), Daejeon 34141, Republic of Korea}

\date{\today}

\bibliographystyle{apsrev4-2}

\begin{abstract}
The invisible axion is a hypothetical particle that arises from the Peccei-Quinn mechanism proposed to resolve the CP problem in quantum chromodynamics, and is considered one of the most favoured candidates for cold dark matter.
Dish antennas can provide a useful scheme for sensitive search for dark matter axions.
The conversion power through axion-photon couplings is proportional to the surface area of the metal plate, rather than the volume of the available magnetic field.
To maximize the effect, we propose an advanced concept of haloscope that involves an array of horn antennae to increase the axion-induced photons and a reflector to focus them onto a photo sensor.
Compared to other proposed schemes, this configuration can significantly improve the experimental sensitivity, especially in the terahertz region.
\end{abstract}

\maketitle


\section{Introduction}

The axion is a hypothetical particle that emerges from the Peccei-Quinn mechanism proposed to solve the strong CP problem of particle physics~\cite{PRL1977PQ,PRL1978Weinberg,PRL1978Wilczek}.
In particular, axions with a mass of sub-eV, called invisible axions, have feeble interactions with ordinary particles, making them suitable for explaining cold dark matter~\cite{PLB1983Wilczek,PLB1983Abbott,PLB1983Dine}.
The representative models of the invisible axion were developed by Kim-Shifman-Vainshtein-Zakharov (KSVZ)~\cite{PRL1979Kim,NPB1980SVZ} and Dine-Fischler-Srednicki-Zhitnitsky (DFSZ)~\cite{YF1980Zhitnitsky,PLB1981DFS}.
Various experimental efforts are underway to look for evidence of dark matter axions~\cite{PRL1987DePanfilis,PRD1990Hagmann,ADMX,CAPP,HAYSTAC,NATURE2022Adair,QUAX,JHEP2020RADES,ORGAN,arXiv2021GrAHal,PRL2022Chang,NATURE2017CAST}.
While the cavity haloscope~\cite{PRD1985Sikivie} has been extensively used for searches in the $\mu$eV region, it becomes impractical at the meV scale where accurate sub-mm alignment is required.
Instead, the dish antenna haloscope (DAH) has been considered as an alternative method in this mass range~\cite{DishAntenna}.
A recent proposal suggests exploiting a self-focusing central parabolic geometry of metal structure, which holds the potential to efficiently search for axions in the terahertz range~\cite{PRL2022Liu}.

Under a static and uniform magnetic background, dark matter axions mix with electric fields that temporally oscillate at the axion frequency, yet are spatially invariant.
Any change of media, e.g., by introducing a dielectric~\cite{MADMAX} or metal plate~\cite{DishAntenna, PRL2022Liu}, causes discontinuities at the interface and the boundary conditions ensure that subsequent electromagnetic waves are emitted perpendicular to the surface.
The DAH employs a conducting dish to induce the axion-photon conversion through such process using a strong magnetic field applied parallel to the dish surface.
The expected power of the photons emitted per unit area $A$ of the metal surface is given by~\cite{DishAntenna}.

\begin{equation}
\label{eq:Parr}
\begin{split}
    \frac{P_{0}}{A} \approx & \left[ \frac{g_{a\gamma\gamma}^{2}\rho_{a}\hbar^{3}}{m_{a}^{2}}\right] \frac{1}{\mu} B_{0}^{2} \\
    \approx & 1.3\times10^{-26}\,{\rm W/m^{2}} \left(\frac{g_{\gamma}}{0.97} \right)^{2} \left(\frac{\rho_a}{0.45\,{\frac{\rm GeV}{\rm cm^3}}} \right) \left(\frac{B_{0}}{\rm 10\,T} \right)^{2},
\end{split}
\end{equation}
where $g_{a\gamma\gamma}$ is the axion-to-photon coupling strength in units of ${\rm GeV^{-1}}$, $\rho_a$ and $m_a$ are the density and mass of the axion dark matter, $\hbar$ is the reduced Planck constant, $B_{0}$ is the strength of the applied magnetic field, $\mu$ is the magnetic permeability, and $g_{\gamma}$ is the model-dependent dimensionless coupling constant with values of $0.97$ and $-0.36$ for the KSVZ and DFSZ models, respectively.
To increase the signal power, DAH experimental designs employ a spherical plate~\cite{DishAntenna} or a plane plate with a parabolic mirror~\cite{PRL2022Liu} to focus the photons generated.

It is worth noting that, unlike the cavity haloscope, the axion-to-photon conversion takes place on the surface of the metal plate, so the surface area, not the volume, within a given magnetic field is important.
Current DAH experiments are therefore not optimally designed for coil-based magnets, which typically generate magnetic fields over the full experimental setup.
Herein, we propose a new DAH concept that utilizes an array of horn antennas to increase the metal surface area in the magnetic field.
The array is aligned in a specific direction so that the converted photons are effectively focused by a parabolic mirror.
This configuration will not only effectively increase the conversion probability for a given magnet setup but also improve the signal power at the focal point.

In Sec.~\ref{sec:horn}, we introduce the concept of the horn array haloscope and calculate the expected axion-to-photon conversion power.
Sec.~\ref{sec:focusing} describes how to improve the direction of propagation of the produced electromagnetic waves to maximize the focusing using a parabolic mirror.
In Sec.~\ref{sec:projection}, the projected sensitivity is calculated and compared with a recently proposed design in terms of projected sensitivity in a practical experimental setup.
Finally, in Sec.~\ref{sec:discussion}, possibilities for further improvement are discussed.

\section{\label{sec:horn}Horn antenna array}

The conversion power of a DAH is proportional to the area of the metal parallel to the applied magnetic field, as described in Eq.~\ref{eq:Parr}.
That is, for a given volume in which a magnetic field exists, introducing as much metal surface area as possible parallel to the magnetic field improves sensitivity.
A trade-off is that the inter plane-spacing becomes smaller and so the searches are limited to higher frequency regions.
This is analogous to the cavity haloscope searches, where multiple small cavities are bundled together~\cite{UCD2001Kinion,MCD2020Yang} or metallic partitions are introduced~\cite{PLB2018Jeong} in order to increase the search frequency while making a full use of the detection volume.
However, in contrast to resonant search, which requires frequency tuning and thus limits the search range, DAH allows for more effective searches at even higher frequencies over a wide band.

There are several ways to effectively increase the surface area in a given magnetic volume.
For a solenoid, multiple layers of concentric metallic hollow cylinders with different radii, a single thin sheet in the shape of a spiral, or many small-radius cylindrical hollows are good examples.
Among these, we consider the third configuration because, similar to multi-cavity systems in cavity search, it offers some advantages over the first two, such as being easily scalable, manufacturable, and replaceable.

\begin{figure}
    \centering
    \hspace*{0.02\linewidth}
    \includegraphics[width=0.35\linewidth]{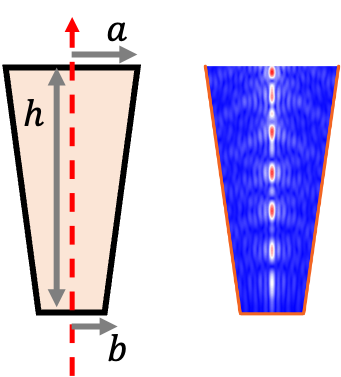} \hspace*{0.02\linewidth}
    \includegraphics[width=0.5\linewidth]{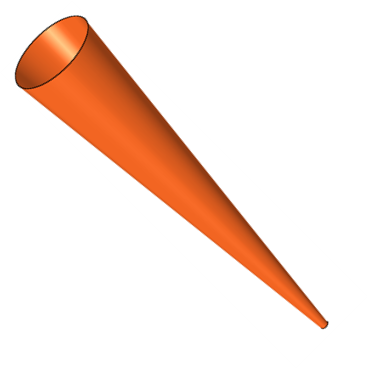}
    \caption{Cross section of a conical horn antenna with its geometric dimensions (left) and its reacted electric field strength on a vertical plane obtained from 2-dimension simulation (center).
    The example horn antenna considered in this work has dimensions of $a=10$\,mm, $b=1$\,mm, and $h=100$\,mm (right).}
    \label{fig:simul_2daxis}
\end{figure}

\begin{figure}
    \centering
    \includegraphics[width=0.95\linewidth]{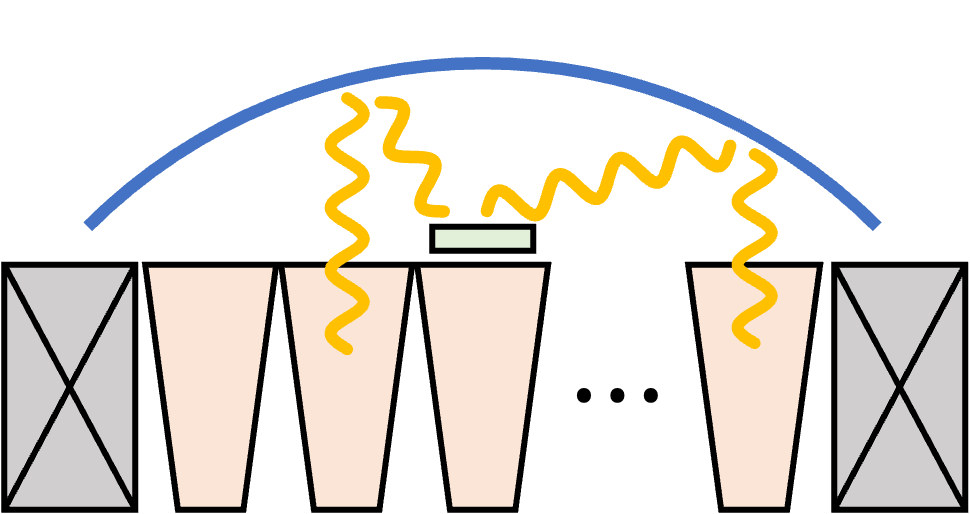}
    \caption{Conceptual sketch of the horn-array haloscope. 
    The array of horn antennas (bisque-color trapezoid) inside the bore of a magnet (gray box) generate photons (yellow curly lines) that are reflected by the parabolic mirror (blue line) and focused onto the photosensor (pale green rectangle) in the middle.
    }
    \label{fig:horn_schematic}
\end{figure}

As the electromagnetic waves generated by dark matter axions propagate perpendicularly to the metal surface, they will bounce around and get trapped inside the hollow cylinders.
An intuitive way to extract/guide them out of the cylinder is to transform the hollow shape into a cone shape, eventually introducing a conical horn antennae, as shown in Fig.~\ref{fig:simul_2daxis}.
The photons exiting along the central axis of each horn antenna can be reflected by a concave parabolic mirror specially designed to focus them onto a photosensor.
The directivity of the photons can be substantially improved by introducing a dielectric lens in front of the opening of the horn antenna, which will be discussed in detail in Sec.~\ref{sec:focusing}.
Many identical horn antennas packed into the solenoid bore, forming a horn-array haloscope, effectively increasing the conversion area.
Figure~\ref{fig:horn_schematic} shows the conceptual schematics of this new haloscope design.

The total power is a linear sum of the power from the individual horn antennas.
The individual power can be estimated by considering the area of a single antenna parallel to the magnetic field.
Assuming the conical shape shown in Fig.~\ref{fig:simul_2daxis} whose dimensions are represented by bottom radius of $b$, top radius of $a$, and height of $h$, the parallel portion of the lateral area of the cone is given by
\begin{equation}
    A_{\rm horn}^{\rm con,\parallel} = A_{\rm horn}^{\rm con} \times \frac{h}{\sqrt{h^{2} + (a - b)^{2}}} = \pi (a + b) h,
\end{equation}
where $A_{\rm horn}^{\rm con}$ is the total lateral area, and $h / \sqrt{h^{2} + (a - b)^{2}}$ is a factor to have the parallel portion to the external magnetic field.
It is noted that a hexagonal opening is more preferable than the circular opening in terms of packing efficiency.
In an infinite limit, the honeycomb structure yields a packing efficiency of unity whereas the circular structure remains at $\pi/2\sqrt{3} \approx 0.91$.
A hexagonal horn antenna with the same dimensions gives a lateral area of
\begin{equation}
    A_{\rm horn}^{\rm hex,\parallel} = 3 (a + b) h.
\end{equation}

For sufficiently high frequencies (short wavelengths), the conversion power from a single conical horn antenna with $a=10$\,mm, $b=1$\,mm, and $h=100$\,mm is

\begin{equation}
\label{eq:Parr_single}
\begin{split}
    P_{\rm single} \approx & \frac{P_{0}}{A} \times A_{\rm horn}^{\parallel} \\
    \approx & 4.5\times10^{-29}\,{\rm W} \left( \frac{B_{0}}{10\,{\rm T}}\right)^{2} \left(\frac{g_{\gamma}}{0.97} \right)^{2} \left( \frac{A_{\rm horn}^{\parallel}}{3460\,{\rm mm^{2}}} \right),
\end{split}
\end{equation}
where we assume $\rho_a=0.45$\,GeV/cm$^3$.

\begin{figure}
    \centering
    \includegraphics[width=0.95\linewidth]{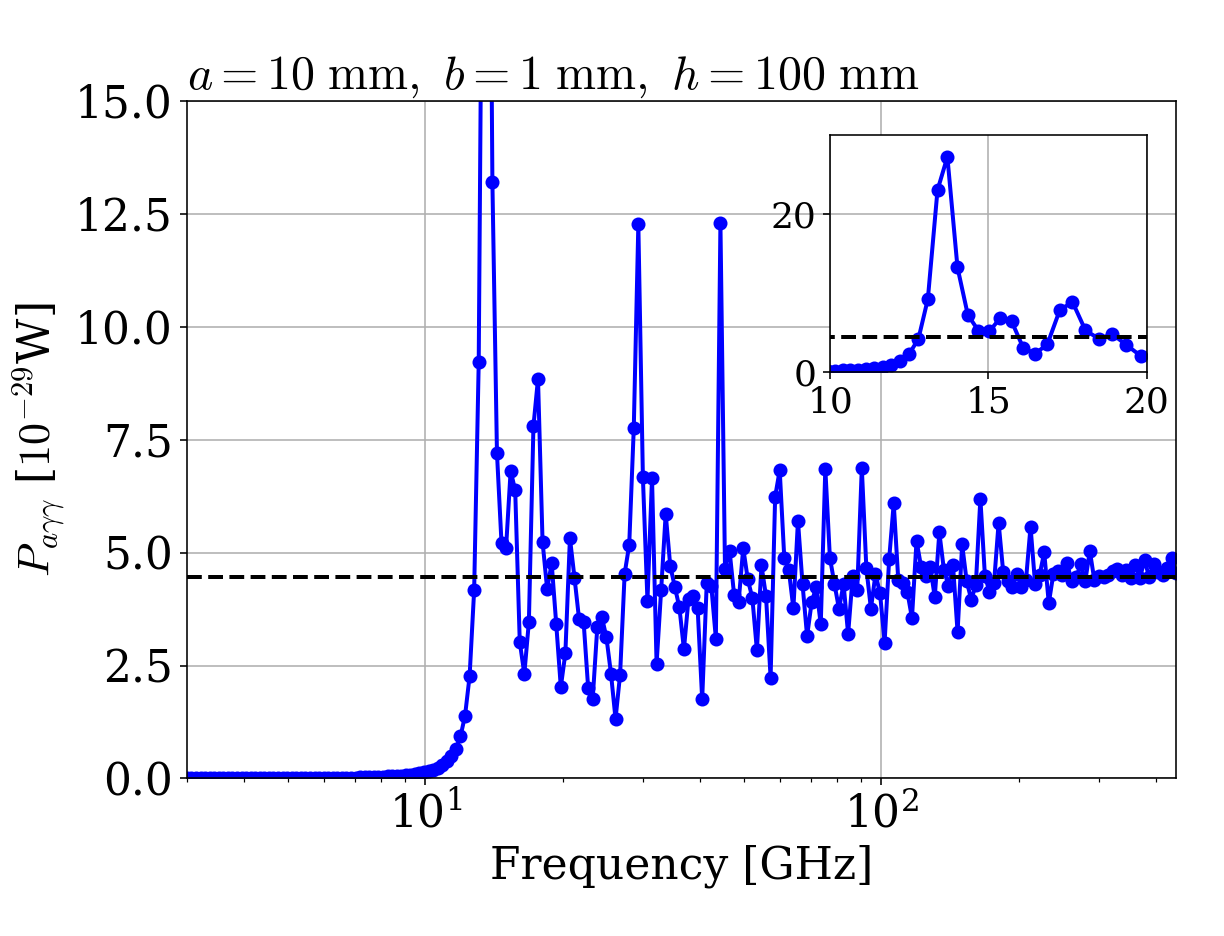}
    \caption{Simulated conversion power for a single horn antenna under a uniform magnetic field. The black dashed line represents the expected conversion power obtained from Eq.~\ref{eq:Parr_single}. 
    A closer view between 10~GHz and 20~GHz is shown in the inset.}
    \label{fig:simul_power_single}
\end{figure}

The conversion power can also be numerically obtained through simulation, for instance, using \texttt{COMSOL Multiphysics}~\cite{COMSOL,arXiv2023DMRadio,JKPS2023Jeong}.
Figure~\ref{fig:simul_power_single} shows the computed conversion power for the horn antenna in a uniform magnetic field as a function of frequency up to 450\,GHz.
The black dashed line represents the estimated power from Eq.~\ref{eq:Parr_single}.
In the low-frequency region, where the wavelength of the photons exceeds the size of the horn antenna, no conversion takes place because the photons are unable to escape the antenna.
As the frequency increases, however, the wavelength becomes smaller than the antenna size, and the numerical calculation eventually converges to the analytical value.
In the intermediate frequency range, resonance occurs intermittently due to the antenna structure, providing a potential opportunity for sensitive narrow-band search.

This new concept offers a notable advantage that the total conversion power scales proportionally with the number of horn antennas configured in an array.
As the number of horn antennas is directly related to the cross-sectional area of the solenoid bore, the total conversion power is enhanced in accordance with the volume of the magnet as
\begin{equation}
\label{eq:Ptotal_hah}
\begin{split}
    P_{\rm total} =& N_{\rm horn} P_{\rm single} \approx \Gamma \left( \frac{R}{b} \right)^{2} P_{\rm single} \\
    \approx & 3.9\times 10^{-24}\,{\rm W} \left( \frac{B_{0}}{10\,{\rm T}}\right)^{2} \left(\frac{g_{\gamma}}{0.97} \right)^{2} \\
    & \left(\frac{\Gamma}{0.75} \right) \left(\frac{R}{0.75\,{\rm m}} \right)^{2} \left(\frac{h}{2.1\,{\rm m}} \right),
\end{split}
\end{equation}
where $\Gamma$ is the packing efficiency, and $R$ is the bore radius of the solenoid.
We still assume $a=10$\,mm and $b=1$\,mm.
Allowing a space of 1\,mm between adjacent horn antennas gives a packing efficiency of about 75\% for a sufficiently large $R$.
This conversion power level is comparable with that from typical cavity haloscopes.

\section{\label{sec:focusing}Radiation focusing}

For effective detection, the emitted photons from the horn antennas are focused on a small area where the photo detector is placed.
In this work, we consider a concave parabolic reflector covering the entire antenna array, as shown in Fig.~\ref{fig:horn_schematic}.
The geometry of the parabola needs to be determined by the desired focal position.
The detection efficiency of this horn array haloscope can be defined as the ratio of the power absorbed by the detector area to the power radiated from the antenna array:
\begin{equation}
    \epsilon = \frac{P_{\rm observed}}{P_{\rm total}}.
\end{equation}
The detection efficiency can also be decomposed into two components: the directivity of the emitted photons from the antenna ($d_{\rm antenna}$) and the focusing efficiency of the paraboloid mirror ($f_{\rm mirror}$) so that
\begin{equation}
    \epsilon \approx d_{\rm antenna} \times f_{\rm mirror}.
\end{equation}

\begin{figure}
    \centering
    \includegraphics[width=0.17\linewidth]{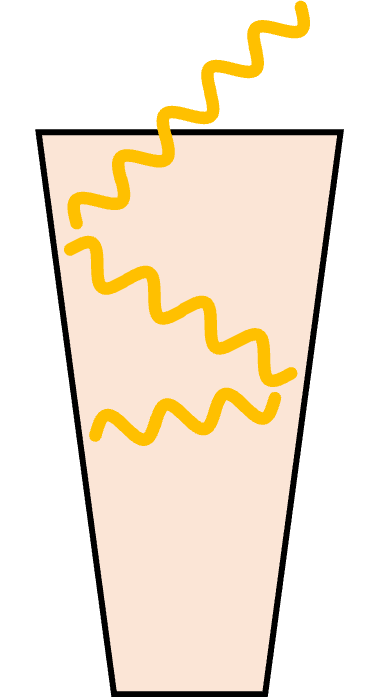}
    \includegraphics[width=0.6\linewidth]{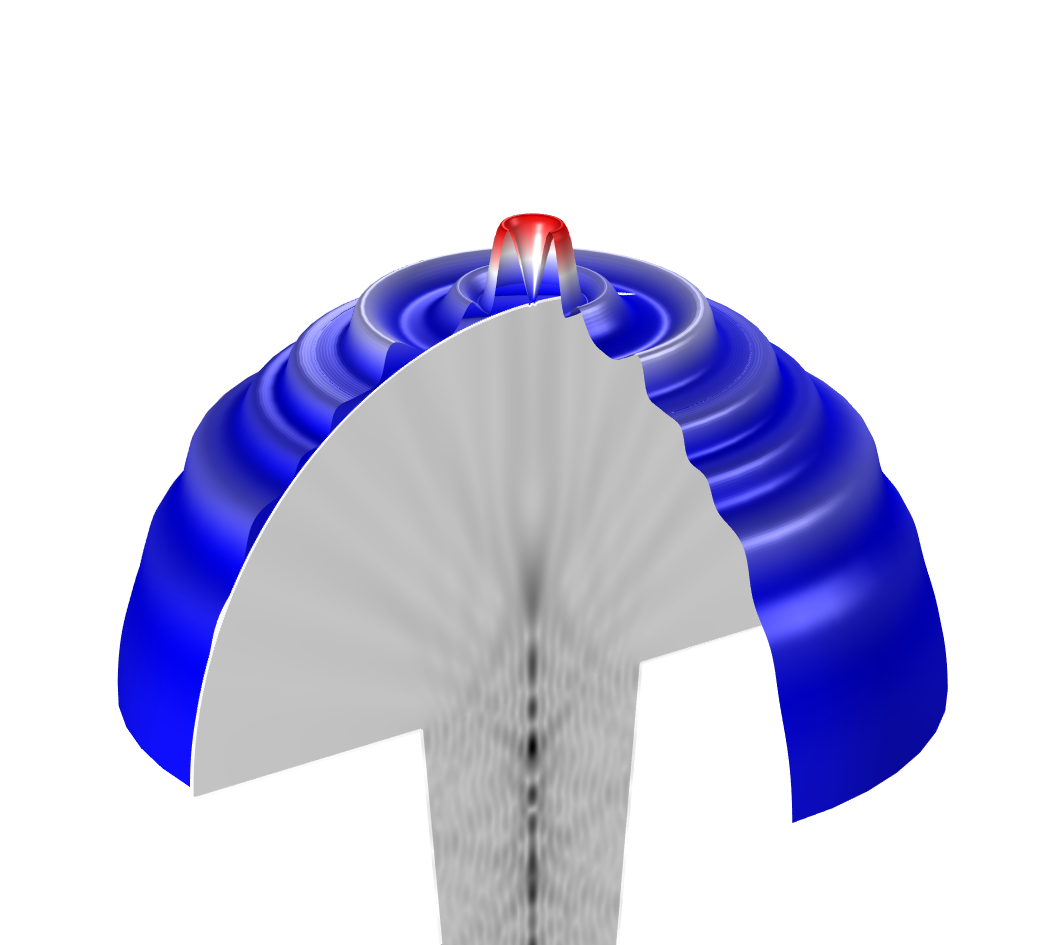}\\
    \includegraphics[width=0.17\linewidth]{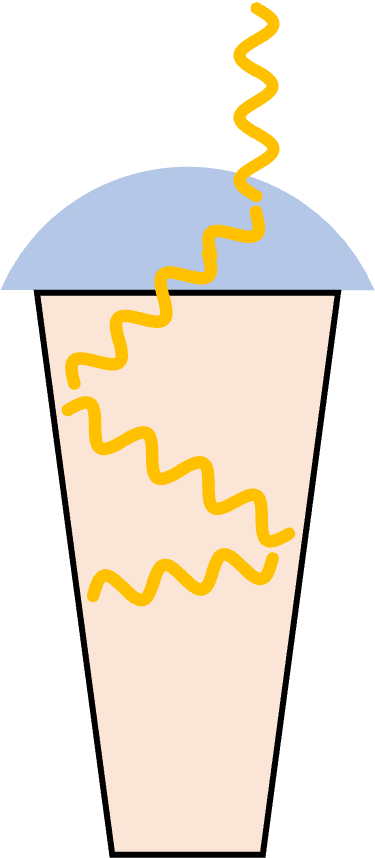}
    \includegraphics[width=0.6\linewidth]{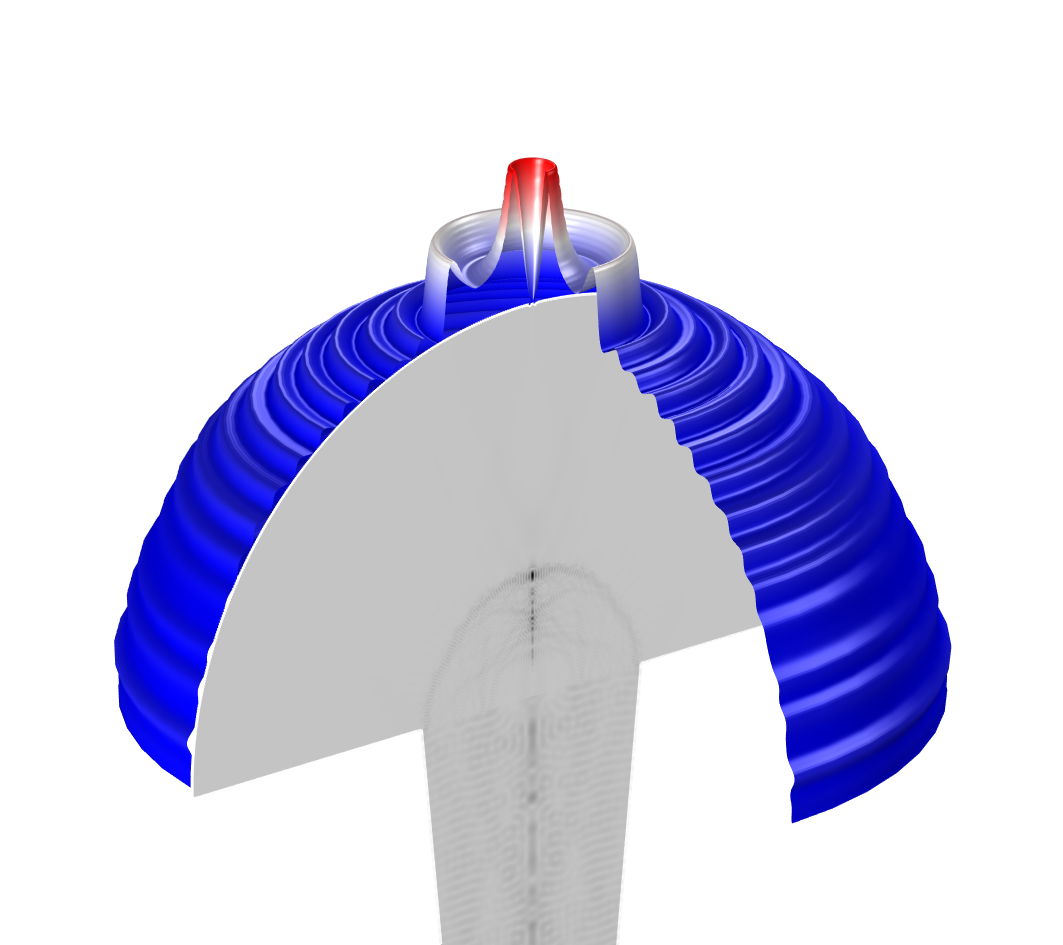}
    \caption{Radiation patterns of a horn antenna without (up) and with (down) a dielectric lens (light blue) at the antenna aperture. The gray-scale color represents the norm of the radiating electric field (the stronger, the darker). The color of the hemispheres represents the relative radiation power (red for strong, blue for weak), while the deformation indicates the radiation pattern.}
    \label{fig:simul_lens}
\end{figure}

A conical or hexagonal horn antenna radiates photons not only in the desired direction but also at different angles.
The directivity can be improved by introducing specially designed internal structures in the antenna~\cite{SciRep2021Eskandari} or a collimator~\cite{IEEE2014Wu,IEEE2017Konstantinidis,SciRep2021Amarasinghe} between the horn antenna and the mirror.
An improvement using a homogeneous dielectric lens placed at the antenna aperture is illustrated in Fig.~\ref{fig:simul_lens}.
It is interesting to note that some particular designs, such as double-ridged horn antennas~\cite{IEEE2020Wu} or axially corrugated horn antennas~\cite{IEEE2016Wang}, can enhance the directivitiy as well as increase the surface area parallel to the magnetic field, further increasing the total conversion power.

The focusing efficiency of the parabolic mirror depends on the design, material and fabrication accuracy of the mirror.
Even if some of the reflected photons are not precisely aimed at the focal point, the focusing efficiency can still be maintained by increasing the receiving area of the detector, for example, by configuring an array of bolometers~\cite{JLTP2018Suzuki}.
With a sufficiently large array of detectors, the overall focusing efficiency can be estimated to be around 0.5 optimistically and 0.1 pessimistically.
The specific optimal design of the horn antenna will be discussed in detail as a follow-up to this study.

\section{\label{sec:projection}Projected sensitivities}

A recently proposal, the Broadband Reflector Experiment for Axion Detection (BREAD), considers a cylindrical metal barrel in a 10-T solenoid for axion-photon conversion and a uniquely designed parabolic mirror in the middle for radiation focusing onto a photosensor~\cite{PRL2022Liu}.
The conversion power follows Eq.~\ref{eq:Parr} and is expected to be: 
\begin{equation}
\begin{split}
    P_{\rm BREAD} \approx& 1.3\times 10^{-25}\,{\rm W} \left( \frac{B_{0}}{10\,{\rm T}}\right)^{2} \left(\frac{g_{\gamma}}{0.97} \right)^{2}\\
    & \left(\frac{R}{0.75\,{\rm m}} \right) \left(\frac{h}{2.1\,{\rm m}} \right).
\end{split}
\end{equation}
It is noted here that in this haloscope design, the power is linearly proportional to the bore radius since it takes place on the side of the barrel which fits the solenoid.
In contrast, the horn array haloscope allows a volume-efficient geometry that makes the conversion power proportional to the square of the bore radius, as seen in Eq.~\ref{eq:Ptotal_hah}.
This innovative design greatly improves the efficiency of the magnet volume usage, yielding a power gain of approximately 30.

\begin{figure}
    \centering
    \includegraphics[width=0.99\linewidth]{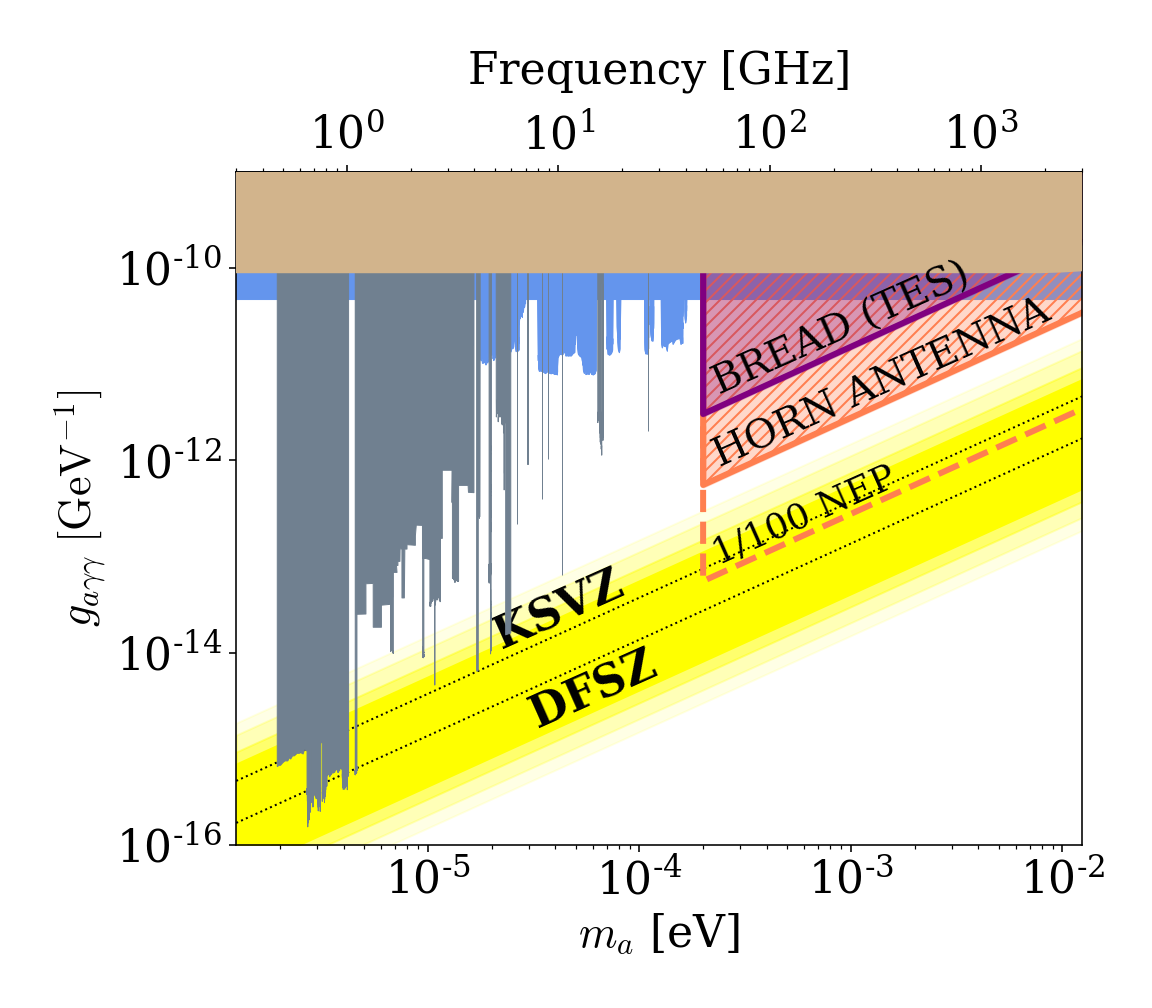}
    \caption{
    Projected sensitivities of the proposed horn-array haloscope based on 3 years of DAQ assuming the current level of NEP (solid orange line) and a 100x improvement (dashed orange line).
    The BREAD sensitivity from the same amount of data is also projected in purple.
    }
    \label{fig:projection}
\end{figure}

The converted photons can be detected by a bolometer such as a transition edge sensor (TES) located at the focal point of the mirror.
Modern TESs available in the terahertz region have noise equivalent power (NEP) of the order of $10^{-19}\,{\rm W/\sqrt{Hz}}$~\cite{JLTP2016Ridder,JAP2011Goldiea,CP2019Kokkoniemi,NATURE2020Lee,JAP2020Paolucci}.
At this noise level, the data acquisition (DAQ) time required to detect photons induced by axions is 
\begin{equation}
\begin{split}
    \Delta t \approx& \left( \frac{{\rm SNR} \times {\rm NEP}}{\epsilon P_{\rm total}} \right)^{2} \\
    =& 300\,{\rm days} \left( \frac{\rm SNR}{5} \right)^{2} \left( \frac{\rm NEP}{2\times10^{-19}\,{\rm W/\sqrt{Hz}}} \right)^{2} \\
    & \left( \frac{0.5}{\epsilon} \right)^{2} \left( \frac{10\,{\rm T}}{B_{0}} \right)^{4} \left(\frac{10 \times 0.97}{g_{\gamma}} \right)^{4} \\
    & \left(\frac{0.75}{\Gamma} \right)^{2} \left(\frac{0.75\,{\rm m}}{R} \right)^{4} \left(\frac{2.1\,{\rm m}}{h} \right)^{2},
    \label{eq:daq_time}
\end{split}
\end{equation}
where $\rm SNR$ is the signal-to-noise ratio desired in an experiment.
Eq.~\ref{eq:daq_time} states that it will take about a year to be sensitive to the dark matter axions with a coupling 10 times stronger than the KSVZ value.
Please note that a DAH is designed for a broadband search, for which no tuning is required, particularly efficient in the terahertz region.
The corresponding projections are shown in Fig.~\ref{fig:projection}.
Compared to the BREAD proposal, it is expected to achieve about 5.5 times improved sensitivity in axion-photon coupling strength.
Moreover, a bolometer with 100-fold lower NEP could probe the KSVZ model with the same amount of data.

\section{\label{sec:discussion}Discussion}

Due to the conical shape of the antenna, there is space available on the other side of the array where another identical array can be configured.
Consequently, it is possible to stack two horn-antenna arrays, facing opposite directions, within a single magnet system.
By incorporating focusing geometries on both the upper and lower sides, it becomes feasible to operate two horn-array haloscopes simultaneously.
An additional structure to combine the signals from the two arrays can be designed to effectively increase the SNR by a factor of 2.

For microwave bolometers utilizing Josephson junctions, experimental setups require a magnetic-free environment.
This is achievable, inspired by the Cassegrain reflector~\cite{IEEE1961Hannan}, by placing a secondary convex mirror at the focal point of the parabolic mirror and forming its focal point at a desired location behind the primary mirror.
This configuration can effectively relocate the detector away from the magnetic field, reducing its impact on measurements.

For experimental schemes designed for broadband searches, the photosensor detects photons originating from a variety of sources over a wide frequency range, making it practically impossible to differentiate the signal from the noise.
However, since our proposed scheme relies on the radiation focusing, defocusing will reduce the detection power in the presence of a desired signal.
Therefore, even in situations where the exact noise level is not known, a variation in power level by focusing or defocusing the reflector indicate the existence of an axion signal.
In addition, incorporating a chopper in front of the detector can be useful for noise calibration particularly when the dominant background source is dark noise, i.e., dark counts by the detector itself.
Such configurations also provide the ability to detect hidden photonic dark matter, as they can identify anomalous power excess without the need to vary the strength of the magnetic field.

\section{Summary}

In this work, we present an improved design of dish antenna haloscope featuring an array of conical horn antennae in a solenoid magnet.
By increasing the metallic surface area within a given magnetic field, this design provides a volume-efficient approach that can substantially increase the axion to photon conversion power. 
A dielectric lens mounted on each horn antenna can improve the directivity of photons into the optimum direction, enabling a parabolic reflector to efficiently focus them onto a photosensor.
Assuming currently available bolometers as photon detectors, the proposed setup can provide an efficient broadband search for axion dark matter in the terahertz region.

\section*{Acknowledgement}
This work was supported by IBS-R017-D1 of the Republic of Korea.

\bibliographystyle{unsrt}

\end{document}